\documentclass[letterpaper,journal]{IEEEtran}
\usepackage{amsmath,amsfonts}
\usepackage{algorithmic}
\usepackage{algorithm}
\usepackage{array}
\usepackage[caption=false,font=normalsize,labelfont=sf,textfont=sf]{subfig}
\usepackage{xcolor} % 确保引入颜色包
\usepackage{color}

\usepackage{textcomp}
\usepackage{stfloats}
\usepackage{url}
\usepackage{verbatim}
\usepackage{subcaption}
\usepackage{graphicx}
\usepackage{subfig}

\usepackage{braket}
\usepackage{cite}
\usepackage[mathscr]{eucal}
\usepackage{amsthm,amsmath,amssymb}
\usepackage{mathrsfs}
\allowdisplaybreaks[1]

\begin{document}

\title{Beamforming Design for Beyond Diagonal RIS-Aided Cell-Free Massive MIMO Systems}

\author{Yizhuo~Li, Jiakang~Zheng,~\IEEEmembership{Member,~IEEE}, Bokai~Xu, Yiyang~Zhu, \\Jiayi~Zhang,~\IEEEmembership{Senior Member,~IEEE}, Dusit~Niyato,~\IEEEmembership{Fellow,~IEEE} and Bo~Ai,~\IEEEmembership{Fellow,~IEEE} 
\vspace{-6mm}

\thanks{This work was supported in part by National Natural Science Foundation of China under Grant 62501042, Grant 62471027, Grant 62221001, and Grant 62341127; in part by China Postdoctoral Science Foundation under Grant BX20250378 and Grant 2024M760195; in part by the Talent Fund of Beijing Jiaotong University under Grant 2024XKRC085; in part by Jiangxi Province Science and Technology development Programme under Grant 20242BCC32016; in part by the Fundamental Research Funds for the Central Universities under Grant 2025JBZX037; and in part by ZTE Industry-University-Institute Cooperation Funds under Grant No. IA20250115003-PO0001. (\emph{Corresponding author: Jiakang Zheng.})}

\thanks{Y. Li, J. Zheng, B. Xu, J. Zhang, and B. Ai are with State Key Laboratory of Advanced Rail Autonomous Operation, and also with School of Electronic and Information Engineering, Beijing Jiaotong University, Beijing 100044, P. R. China (e-mail: {jiakangzheng, jiayizhang, boai}@bjtu.edu.cn).
}
\thanks{Y. Zhu is with the School of Electrical and Electronics Engineering, Nanyang Technological University, Singapore 639798.}
\thanks{D. Niyato is with the College of Computing and Data Science, Nanyang Technological University, Singapore 639798 (e-mail: dniyato@ntu.edu.sg).}

}

\maketitle

\begin{abstract}Reconfigurable intelligent surface (RIS)-aided cell-free (CF) massive multiple-input multiple-output (mMIMO) is a promising technology for further improving spectral efficiency (SE) with low cost and power consumption. However, conventional RIS has inevitable limitations due to its diagonal scattering matrix. In contrast, beyond-diagonal RIS (BD-RIS) has gained great attention. This correspondence focuses on integrating a hybrid transmitting and reflecting BD-RIS into CF mMIMO systems to enhance coverage and spatial multiplexing. This requires completing the beamforming design under the transmit power constraints and unitary constraints of the BD-RIS, by optimizing active and passive beamformer simultaneously. To tackle this issue, we introduce an alternating optimization algorithm that decomposes it using fractional programming and solves the subproblems alternatively. Moreover, to address the challenge introduced by the unitary constraint on the beamforming matrix of the BD-RIS, we propose a Riemannian limited-memory Broyden-Fletcher-Goldfarb-Shanno (R-L-BFGS) algorithm  to solve the problem optimally.
Simulation results show that our algorithm achieves faster convergence and finds higher-quality solutions compared with baselines, while also demonstrating a favorable performance–complexity trade-off.

\end{abstract}

\begin{IEEEkeywords}Beyond-diagonal RIS, cell-free massive MIMO, spectral efficiency, beamforming design.

\end{IEEEkeywords}

\vspace{-2mm}
\section{Introduction}
 The upcoming sixth generation (6G) network is anticipated to play a key role in many areas of society, industry and daily life in the future, requiring extremely high communication standards in terms of capacity, latency, reliability and intelligence\cite{9040431, 11058968}. In the future, cell-free (CF) massive multiple-input multiple-output (mMIMO) systems are expected to replace traditional cellular systems in empowering the upcoming 6G\cite{3,r1,r2,zhengjiak}. By distributing access points (APs) across a wide area and connecting them to a central processing unit (CPU), CF mMIMO systems can offer high data rates, low latency, and efficient resource utilization. Moreover, reconfigurable intelligent surfaces (RISs) have deployed as an economical solution in the CF mMIMO system\cite{zhu2025joint,mazhangfa,bokai}.

RIS is an advanced technology that enhances signal propagation and quality by using a large array of passive, low-cost reflective elements to dynamically adjust the phase and amplitude of incident electromagnetic waves. However, traditional RIS inevitably has limitations in complex scenarios because the scattering matrix must be diagonal. To address the limitations of RIS, the beyond-diagonal RIS (BD-RIS) has been proposed as a novel evolution\cite{1}. Unlike traditional RIS,
BD-RIS exhibits a non-diagonal scattering matrix. Depending on the interconnection schemes among the elements, BD-RIS can be classified into: single-connected (SC), fully-connected (FC), group-connected (GC), tree-connected, band-connected \cite{gra} and stem-connected \cite{stem} architectures. According to the arrangement of elements for each architecture, BD-RIS has three modes: reflecting mode, hybrid transmitting and reflecting mode and multi-sector mode\cite{1}.

The recent researchs have delved into BD-RIS-aided systems. The performance of BD-RISs with different modes/architectures in multiuser multiple input single
output (MU-MISO) system had been studied comprehensively in\cite{1}.
The authors in \cite{STAR} conducted an in-depth study of the performance of STAR-RIS in CF mMIMO systems.
Although STAR-RIS introduced the concept of hybrid transmission and reflection, it is fundamentally a single-connected (SC) architecture.
This authors in \cite{SWIFT} evaluates the performance of a BD-RIS-aided CF mMIMO for simultaneous wireless information and power transfer (SWIPT) under practical channel and energy harvesting assumptions.
And the authors of \cite{10839400} consider a BD-RIS-aided multi-user system and use iterative closed-form solutions to alternately optimize active and passive beamforming matrices.
The aforementioned works have showcased the outstanding performance of BD-RIS in communication systems. However, to the best of our knowledge, the performance of hybrid BD-RIS in more architecture in CF mMIMO systems has not been studied.

In this correspondence, we investigate the integration of a hybrid transmitting and reflecting BD-RIS into CF mMIMO systems to enhance both coverage and spatial multiplexing. The main contributions are summarized as follows:
\begin{itemize}
    \item {We develop a unified beamforming design framework that enables a comprehensive performance comparison among various BD-RIS architectures, including SC, GC, and FC.}
    \item {To effectively solve the non-convex optimization problem, we propose a Riemannian limited-memory Broyden–Fletcher–Goldfarb–Shanno (R-L-BFGS) algorithm, which offers a robust and efficient solution.}
    \item {Simulation results demonstrate that the proposed R-L-BFGS algorithm achieves faster convergence and attains higher-quality solutions compared with baselines.}
\end{itemize}

\vspace{-4mm}
\section{SYSTEM MODEL}

We focus on a BD-RIS-aided CF mMIMO system. The system consists of ${L}$ APs, ${K}$ UEs, and one BD-RIS. Each AP and UE consist of ${N}$ antennas and one antenna, respectively. The BD-RIS has $M$ cells, each consisting of two back-to-back antennas interconnected via reconfigurable admittance components\cite{1}. The BD-RIS serves $K_r$ reflective and $K_t$ transmissive UEs, with a total of $K=K_r+K_t$. Let $\mathcal{L}=\{1,\ldots,L\}$, $\mathcal{K}_{r}=\{1,\ldots,K_{r}\}$, $\mathcal{K}_{t}=\{1,\ldots,K_{t}\}$ and $\mathcal{K}=\{1,\ldots,K\}$ represent the index sets for APs, reflective UEs, transmissive UEs and all UEs respectively. We assume perfect channel state information (CSI) is available at the APs. 
Then, the channel from the ${l}$-th AP to the ${k}$-th UE is{{\setlength{\abovedisplayskip}{9pt}
 \setlength{\belowdisplayskip}{9pt}
\begin{align}
\mathbf{h}_{l,k}^{\mathrm{H}}=\mathbf{f}_{k}^{\mathrm{H}}\mathbf{\Theta}_{i}\mathbf{G}_{l}+\mathbf{h}_{l,k,d}^{\mathrm{H}}, \forall i\in\{\mathrm{t},\mathrm{r}\},\forall k\in\mathcal{K}_{i},\label{4}
\end{align}}
\parskip=0pt
where $\mathbf{f}_{k}\in\mathbb{C}^{M}$, $\mathbf{h}_{l,k,d}\in\mathbb{C}^{N}$ and $\mathbf{G}_{l}\in\mathbb{C}^{M\times N}$ represent the frequency-domain channel from the BD-RIS to the UE \textit{k}, from the AP ${l}$ to the UE \textit{k} and from the AP ${l}$ to the BD-RIS, respectively. 
In the GC BD-RIS, the $M$ cells are partitioned into $G$ groups, denoted by $\mathcal{G}=\{1,\ldots,G\}$. Each group has an equal size $\bar{M}=M/G$. Other BD-RIS scenarios can be handled by adjusting $G$ accordingly.. We consider that each group has an equal size, denoted as $\bar{M}=M/G$.
In the hybrid BD-RIS, the scattering matrix $\mathbf{\Theta}$ is divided into $\mathbf{\Theta}_\mathrm{r}^{}$ and $\mathbf{\Theta}_\mathrm{t}^{}$. $\mathbf{\Theta}_\mathrm{r}^{}$ and $\mathbf{\Theta}_\mathrm{t}^{}$ are block diagonal matrices, expressed as: $\mathbf{\Theta}_{\mathrm{r}}= \mathrm{blkdiag}(\mathbf{\Theta}_{_{\mathrm{r},1}},\ldots,\mathbf{\Theta}_{_{\mathrm{r},G}})$, $\mathbf{\Theta}_{\mathrm{t}}=\mathrm{blkdiag}(\mathbf{\Theta}_{_{\mathrm{t},1}},\ldots,\mathbf{\Theta}_{_{\mathrm{t},G}})$,
where $\mathbf{\Theta}_{\mathrm{t},g},\mathbf{\Theta}_{\mathrm{r},g}\in\mathbb{C}^{\bar{M}\times\bar{M}},\forall g\in\mathcal{G}$ satisfy the constraint: 
$\mathbf{\Theta}_{\mathrm{r},g}^\mathrm{H}\mathbf{\Theta}_{\mathrm{r},g}+\mathbf{\Theta}_{\mathrm{t},g}^\mathrm{H}\mathbf{\Theta}_{\mathrm{t},g}=\mathbf{I}_{\bar{M}}.$ 
Thus the received signal by the $k$-th UE is
% {\setlength{\abovedisplayskip}{9pt}
%  \setlength{\belowdisplayskip}{9pt}
% \begin{align}
% y_{k} =\mathbf{h}_{k}^{\mathrm{H}}\mathbf{w}_{k}s_{k}+\sum_{j=1,j\neq k}^{K}\mathbf{h}_{k}^{\mathrm{H}}\mathbf{w}_{j}s_{j}+n_{k},
% \end{align}}
\begin{align}
{y_k} = {\bf{h}}_k^{\rm{H}}{{\bf{w}}_k}{s_k} + \sum\nolimits_{j = 1,j \ne k}^K {{\bf{h}}_k^{\rm{H}}{{\bf{w}}_j}{s_j} + {n_k}} ,
\end{align}
where $n_{k} \sim \mathcal{C N}\left(0, \sigma_{k}^{2}\right)$ denotes the additive white Gaussian noise, $\mathbf{s}_{k}$, $\forall k \in \mathcal{K}$ denotes the transmitted symbol to the ${k}$-th UE, $\mathbf{w}_{l, k} \in \mathbb{C}^{N}$ is the active beamformer for UE ${k}$ at the ${l}$-th AP. 
And we defined $\mathbf{h}_{k}=[\mathbf{h}_{1,k}^{\mathrm{T}},\ldots,\mathbf{h}_{L,k}^{\mathrm{T}}]^{\mathrm{T}}\in\mathbb{C}^{LN}$ and $\mathbf{w}_{k}=[\mathbf{w}_{1,k}^{\mathrm{T}},\ldots,\mathbf{w}_{L,k}^{\mathrm{T}}]^{\mathrm{T}}\in\mathbb{C}^{LN}$.
Then, the signal-to-interference-plus-noise ratio for each UE is derived as
\begin{align}
{\gamma _k} = |{\bf{h}}_k^{\rm{H}}{{\bf{w}}_k}{|^2}/\left( {\sum\nolimits_{j \in {\cal K},j \ne k} {{{\left| {{\bf{h}}_k^{\rm{H}}{{\bf{w}}_j}} \right|}^2} + \sigma _k^2} } \right),\forall k \in {\cal K}.
\end{align}
Moreover, the SE maximization problem is
formulated as{\setlength{\abovedisplayskip}{9pt}
 \setlength{\belowdisplayskip}{9pt}
\begin{subequations}\begin{align}
\mathcal{P}^{\mathrm{o}}:\operatorname*{max}_{\mathbf{w},\mathbf{\Theta}_{\mathrm{t}},\mathbf{\Theta}_{\mathrm{r}}}&\text{sum-SE}
=\sum\nolimits_{k\in\mathcal{K}}\log_{2}(1+\gamma_{k}) \label{10a}\\
\mathrm{s.t.}\quad C_{1}&:\mathbf{\Theta}_{\mathrm{r,g}}^{\mathrm{H}}\mathbf{\Theta}_{\mathrm{r,g}}+\mathbf{\Theta}_{\mathrm{t,g}}^{\mathrm{H}}\mathbf{\Theta}_{\mathrm{t,g}}=\mathbf{I}_{\bar{M}},\forall\:g\in\mathcal{G}, \label{10b}\\
C_{2}&:\sum\nolimits_{k=1}^{K}\parallel \mathbf w_{l,k}\parallel^{2}\leq P_{l,\mathrm{max}}, \label{10c}
\end{align}\end{subequations}}
where ${P} _{l,\mathrm{max}}$ is the maximum transmit power of the ${l}$-th AP and $\mathbf{w}=[\mathbf{w}_{1}^{\mathrm{T}},\ldots,\mathbf{w}_{K}^{\mathrm{T}}]^{\mathrm{T}}\in\mathbb{C}^{KLN}$ is defined for simplicity.

\vspace{-4mm}
\section{BEAMFORMING DESIGN}
\subsection{Overview of the Beamforming Design Framework}
Utilizing the Lagrangian dual transform and quadratic
transform\cite{5}, the objective sum-SE can be restated as follows:
\begin{align}
&f_{\tau}(\mathbf{w},\mathbf{\Theta}_{\mathrm{t}},\mathbf{\Theta}_{\mathrm{r}},\boldsymbol{\rho},\boldsymbol{\tau}) =
\sum\nolimits_{k\in\mathcal{K}} \Big( 
\log_2(1+\rho_k) - \rho_k \notag\\
+ &2\sqrt{1+\rho_k}\, \Re \{\tau_k^* \mathbf{h}_k^{\mathrm{H}} \mathbf{w}_k\} 
- |\tau_k|^2 \sum_{j\in\mathcal{K}} (|\mathbf{h}_k^{\mathrm{H}} \mathbf{w}_j|^2 + \sigma_k^2)
\Big), \label{100}
\end{align}
where $\boldsymbol{\rho}\triangleq[\rho_{1},\ldots,\rho_{K}]^{\mathrm{T}}\in\mathbb{R}^{K}, \boldsymbol{\tau}\triangleq[\tau_1,\ldots,\tau_K]^\mathrm{T}\in\mathbb{C}^K$. 
Given fixed $(\mathbf{w},\boldsymbol{\Theta}_t,\boldsymbol{\Theta}_\mathrm{r},\boldsymbol{\rho}$ or $\boldsymbol{\tau})$, the optimal $\boldsymbol{\tau}$ or $\boldsymbol{\rho}$ can be obtained
by solving $\frac{\partial f_{\tau}}{\partial\tau_{k}}=0$ or $\frac{\partial f_{\tau}}{\partial\rho_{k}}=0$ for $\forall k\in\mathcal{K}$. The solution can be written as
{\setlength{\abovedisplayskip}{9pt}
 \setlength{\belowdisplayskip}{9pt}
\begin{align}
\rho_k^\star=\gamma_k,
\tau_{k}^{\star}=\frac{\sqrt{1+\rho_{k}} \mathbf{h}_{k}^{\mathrm{H}}\mathbf{w}_{k}}{\sum_{j\in\mathcal{K}}| \mathbf{h}_{k}^{\mathrm{H}} \mathbf{w}_{j} |^{2}+\sigma_{k}^{2}},\forall k\in\mathcal{K}.\label{15}
\end{align}
}
\subsection{Active Beamforming: Solve $\mathbf{w}$}
Given $(\boldsymbol\Theta_{\mathrm{t}}^{\star},\boldsymbol\Theta_\mathrm{r}^{\star},\boldsymbol\rho^{\star},\boldsymbol\tau^{*})$,
We first define $\mathbf{a}_{}=\sum_{k\in\mathcal{K}}\mathbf{h}_{k}\tau_{k}\tau_{k}^{*}\mathbf{h}_{k}^{\mathrm{H}},
\mathbf{A}=\mathbf{I}_K\otimes\mathbf{a}, \mathbf{v}_{k}=\mathbf{h}_{k}\tau_{k}.$
Then, we can extract the equations in (\ref{100}) associated with $\mathbf{w}$ as
\begin{equation}g_1\left(\mathbf{w}\right)=-\mathbf{w}^\mathrm{H}\mathbf{A}\mathbf{w}+\mathfrak{R}\left\{2\mathbf{v}^\mathrm{H}\mathbf{w}\right\}-Y,\end{equation}
 where 
$Y=\sum_{k\in\mathcal{K}}\tau_{k}^\mathrm{H}\sigma_{k}\tau_{k}$, $\mathbf{v}=[\mathbf{v}_{1}^\mathrm{T},\mathbf{v}_{2}^\mathrm{T},\ldots,\mathbf{v}_{K}^\mathrm{T}]^\mathrm{T}$. Therefore, the problem can be further simplified as
\begin{equation}\begin{aligned}
\hat{\mathcal{P}}:&\min_{\mathbf{w}}\ \ g_{2}(\mathbf{w})=\mathbf{w}^\mathrm{H}\mathbf{A}\mathbf{w}-\mathfrak{R}\left\{2\mathbf{v}^{\mathrm{H}}\mathbf{w}\right\}\\
&\ \mathrm{s.t.}\quad\mathbf{w}^{\mathrm{H}}\mathbf{D}_{l}\mathbf{w}\leq P_{l,\mathrm{max}},\forall l\in\mathcal{L},\label{00}\end{aligned}\end{equation}
where $\mathbf{D}_{l}=\mathbf{I}_{K} \otimes\left\{\left(\mathbf{e}_{l} \mathbf{e}_{l}^{\mathrm{H}}\right) \otimes \mathbf{I}_{M}\right\}$ with $\mathbf{e}_l\in\mathbb{R}^L$.
Since $\mathbf{A}$ and $\mathbf{D}_{l}$ ($\forall l\in\mathcal{L}$) are all positive semidefinite, the
simplified subproblem $\hat{\mathcal{P}}$ is a standard quadratically constrained quadratic program problem.
To reduce the complexity, we exploit the \textit{primal-dual
subgradient (PDS)} method in \cite{3} to obtain $\mathbf{w}^\star$.\parskip=0pt

\subsection{Passive Beamforming: Solve $\{$$\mathbf{\Theta}_\mathrm{t}, \mathbf{\Theta}_\mathrm{r}$$\}$ }
When $\mathbf{w}$, $\boldsymbol\rho$ and $\boldsymbol\tau$ are fixed, we can first define
\parskip=0pt
\begin{equation}\begin{aligned}
&\eta_{k}=\sqrt{1+\rho_{k}} \tau_{k}, \mathbf{g}_{k} =\sum\nolimits_{l=1}^{L}\mathbf{G}_{l}\mathbf{w}_{l,k}, \\&a_{k,j}=\sum\nolimits_{l=1}^{L}\mathbf{h}_{l,k,d}^\mathrm{H}\mathbf{w}_{l,j}, \forall k \in \mathcal{K},\forall j \in \mathcal{K},
\end{aligned}\end{equation}\parskip=0pt
and then the sub-objective function with
respect to passive beamformer $\mathbf{\Theta}_\mathrm{t}$ and $\mathbf{\Theta}_\mathrm{r}$ can be written as
\begin{equation}\begin{aligned}
 \sum_{i\in\{\mathrm{t},\mathrm{r}\}}&\sum_{k\in\mathcal{K}_{i}}(2\Re\{\mathbf{f}_{k}^{\mathrm{H}}\mathbf\Theta_{i}\mathbf{t}_{k}\}-\left| \tau_{k}\right|^{2}\sum_{j\in\mathcal{K}}\left|\mathbf{f}_{k}^{\mathrm{H}}\mathbf\Theta_{i}\mathbf{g}_{j}\right|^{2}),\label{18}
\end{aligned}\end{equation}

where $\mathbf{b}_{k}\triangleq\left|\tau_{k}\right|^{2}\sum_{j\in\mathcal{K}}a_{k,j} ^{*}\mathbf{g}_{j}, \mathbf{t}_{k}=\eta_{k}^{*}\mathbf{g}_{k}-\mathbf{b}_k,\forall k \in \mathcal{K}.$ And (\ref{18}) can be further reformulated as
\begin{equation}\begin{aligned}
\sum\nolimits_{i\in \{\text{t},\text{r}\}}&{\left( 2\Re \left\{ \mathrm{Tr}\left( {{\mathbf\Theta }_{i}}{{\mathbf{A}}_{i}} \right) \right\}-\mathrm{Tr}( {{\mathbf\Theta }_{i}}\mathbf{B\Theta }_{i}^{\mathrm{H}}{{\mathbf{C}}_{i}})\right), }
\label{19}
\end{aligned}
\end{equation}
where
$\mathbf{A}_{i}\triangleq\sum_{k\in\mathcal{K}_{i}}\mathbf{t}_{k}\mathbf{f}_{k}^{\mathrm{H}}\in\mathbb{C}^{M\times M}, \mathbf{B}\triangleq\sum_{j\in\mathcal{K}}\mathbf{g}_{j}\mathbf{g}_{j}^{\mathrm{H}}\in\mathbb{C}^{M\times M},
\mathbf{C}_{i}\triangleq\sum_{k\in\mathcal{K}_{i}} | \tau_{k} |^{2} \mathbf{f}_{k}\mathbf{f}_{k}^{\mathrm{H}}\in\mathbb{C}^{M\times M},\forall i\in\{\mathrm{t},\mathrm{r}\}.$
However, the quadratic terms in  (\ref{19}) lead to interactions among different groups of $\boldsymbol\Theta_{\mathrm{t/r},g}$. To address this, we aim to isolate one specific pair $\boldsymbol\Theta_{\mathrm{t},g}$ and $\boldsymbol\Theta_{\mathrm{r},g}$ for a given group ${g}$ while keeping the other pairs fixed, focusing on optimizing the selected pair.
For this purpose, we first reformulate objective in (\ref{19}) on a group-by-group basis as follows:\parskip=0pt
\begin{equation}
\begin{aligned} 
&\!\!\!\! \sum\limits_{i \in \{ {\rm{t}},{\rm{r}}\} } \!\!( 2\Re \left\{ {{\rm{Tr}}({{\bf{\Theta }}_i}{{\bf{A}}_i})} \right\} \!\!-\!\! {\rm{Tr}}({{\bf{\Theta }}_i}{\bf{B\Theta }}_i^{\rm{H}}{{\bf{C}}_{\rm{i}}})) \!=\!\!\! \sum\limits_{i \in \{ {\rm{t}},{\rm{r}}\} }\!\! {(2} \sum\limits_{q = 1}^G {} \\
 &\!\!\!\!\!\!\times \! \Re \{ {\rm{Tr}}({{\bf{\Theta }}_{i,q}}{{\bf{A}}_{i,q}})\}  \!-\! {\rm{Tr}}(\sum\limits_{p = 1}^G \!{{{\bf{\Theta }}_{i,p}}} \!\! \sum\limits_{q = 1}^G \!{{{\bf{B}}_{p,q}}} {\bf{\Theta }}_{i,q}^{\rm{H}}{{\bf{C}}_{i,q,p}})),
\end{aligned}
\end{equation}
\parskip=0pt where \parskip=0pt
\begin{equation}
\begin{aligned} 
&\mathbf{A}_{i,q}=[\mathbf{A}_{i}]_{(q-1)\bar{M}+1:q\bar{M},(q-1)\bar{M}+1:q\bar{M}},\\&\mathbf{B}_{p,q}=[\mathbf{B}]_{(p-1)\bar{M}+1:p\bar{M},(q-1)\bar{M}+1:q\bar{M}}, \\&\mathbf{C}_{i,q,p}=[\mathbf{C}_{i}]_{(q-1)\bar{M}+1:q\bar{M},(p-1)\bar{M}+1:p\bar{M}},\forall p,q\in\mathcal{G}.
\end{aligned}
\end{equation}
When focusing on a single pair $\boldsymbol\Theta_{\mathrm{t},g}$ and $\boldsymbol\Theta_{\mathrm{r},g}$, the resulting sub-objective function takes the following form:
\begin{equation}\begin{aligned}
 f_{g}(\boldsymbol\Theta_{t,g},\boldsymbol\Theta_{t,g})=\sum_{i\in(\mathrm{t,r})}\Bigl(\mathrm{Tr}(\boldsymbol\Theta_{i,g}\mathbf{B}_{g,g} \boldsymbol\Theta_{i,g}^{\mathrm{H}}\mathbf{C}_{i,g,g})\\-2\Re\{\mathrm{Tr}(\boldsymbol\Theta_{i,g}\underbrace{(\mathbf{A}_{i,g}-\sum_{p\neq g}\mathbf{B}_{g,p}\boldsymbol\Theta_{i,p}^{\mathrm{H}}\mathbf{C}_{i,p,g})}_{\mathbf{X}_{i,g}})\}\Bigr).
\end{aligned}\end{equation}
Define matrices
$\mathbf{\Theta}_{g}\triangleq[\mathbf{\Theta}_{t,g}^{\mathrm{H}},\mathbf{\Theta}_{r,g}^{\mathrm{H}}]^{\mathrm{H}}$, $\mathbf{X}_{g}\triangleq[\mathbf{X}_{t,g},\mathbf{X}_{r,g}]$, and $\mathbf{C}_{g}\triangleq\mathrm{blkdiag}(\mathbf{C}_{t,g,g},\mathbf{C}_{r,g,g})$. 
We can then express the sub-problem related to $\mathbf{\Theta}_{g}$ as follows:
\begin{subequations}\begin{align}\min_{\boldsymbol{\Theta_{g}}}& \quad\tilde{f}_{g}(\boldsymbol{\Theta}_{g})=\mathrm{Tr}(\boldsymbol{\Theta}_{g}\mathbf{B}_{g,g}\mathbf{\Theta}_{g}^{\mathrm{H}}\mathbf{C}_{g}-2\Re\{\boldsymbol{\Theta}_{g}\boldsymbol{X}_{g}\})\\&\mathrm{s.t.} \quad\boldsymbol{\Theta}_{g}^{\mathrm{H}}\boldsymbol{\Theta}_{g}=\mathbf{I}_{\bar{M}},\forall g\in\mathcal{G}.\label{23b}\end{align}\end{subequations}
To address the non-convex constraint \eqref{23b}, we introduce manifold algorithm\cite{7}. (\ref{23b}) defines an
Stiefel manifold:
\begin{equation}\begin{aligned}
\mathcal{M}_g=\{\boldsymbol\Theta_g\in\mathbb{C}^{2\bar{M}\times\bar{M}}:\boldsymbol\Theta_g^\mathrm{H}\boldsymbol\Theta_g=\mathbf{I}_{\bar{M}}\},\forall g\in\mathcal{G}.\end{aligned}\end{equation}
And the tangent space for $\mathcal{M}_g$ at $\boldsymbol\Theta_{g,k}$ is given by
\begin{equation}{{T}}_{\boldsymbol\Theta_{g,k}}\mathcal{M}_g\!=\!\{\mathbf{T}_g\!\in\!\mathbb{C}^{2\bar{M}\times\bar{M}}\!:\!\Re\{\mathbf({\boldsymbol\Theta_{g,k}})^\mathrm{H}\mathbf{T}_g\}\!=\!\mathbf{0}_{\bar{M}}\},\forall g\in\mathcal{G}.\end{equation}
The Riemannian gradient can be calculated by projecting
the Euclidean gradient onto the tangent space\cite{7}:
\begin{equation}\begin{aligned}
\mathrm{grad}\,\tilde{f}_g(\mathbf{\Theta}_{g,k})
\!\!=\!\!\triangledown\tilde{f}_{g}(\boldsymbol{\Theta}_{g,k})\!-\!{\mathbf\Theta}_{g,k}\mathrm{chdiag}({\mathbf\Theta}_{g,k})^\mathrm{H}\triangledown\tilde{f}_{g}(\boldsymbol{\Theta}_{g,k}),\label{27}\end{aligned}\end{equation}
 where $\triangledown\tilde{f}_g(\boldsymbol{\Theta}_{g,k})=2\mathbf{C}_g\boldsymbol{\Theta}_{g,k}\mathbf{B}_{g,g}-2\mathbf{X}_g^\mathrm{H},\forall g\in\mathcal{G}$ and
$\mathrm{chdiag(\cdot)}$ chooses all diagonal elements of a matrix to construct a diagonal matrix.
The iterative update is given by 
$
\mathbf{\Theta}_{g,k+1} = R_{\mathbf{\Theta}_{g,k}}(\alpha_k \mathbf{\eta}_k),$
where $\mathbf{\Theta}_{g,k}$ is the current iterate, $\mathbf{\eta}_k \in T_{\mathbf{\Theta}_{g,k}}\mathcal{M}_g$ is the search direction in the tangent space, $\alpha_k$ is the step length which can be searched by backtracking
algorithms\cite{7}, and $R_{\mathbf{\Theta}_{g,k}}(\cdot)$ is a retraction mapping a tangent vector back onto the manifold.
\begin{algorithm} 
\caption{R-L-BFGS Two-Loop Recursion}  
\label{alg:two_loop}
\begin{algorithmic}[1] % [1] 表示每行都显示行号
 \REQUIRE Current gradient $\mathbf{g}_k$, history buffers $S=\{\mathbf{s}_i\}_{i=k-m+1}^{k}$,$Y=\{\mathbf{y}_i\}_{i=k-m+1}^{k}$, $P=\{\rho_i\}_{i=k-m+1}^{k}$.
 \ENSURE New search direction $\mathbf{\eta}_{k+1}=-\mathcal{B}_{k+1}\mathbf{g}_{k+1}$.
    
    \STATE $\mathbf{r} = \mathbf{g}_{k+1}$

    \FOR{$i = k: {k-m+1}$}
        \STATE $\alpha_i = \rho_i\,\langle\mathbf{s}_i,\mathbf{r}\rangle$
        \STATE $\mathbf{r}  = \mathbf{r} - \alpha_i\mathbf{y}_i$
    \ENDFOR
   \STATE $\mathbf{r} = \mathcal{B}_k^0\mathbf{r}$
     \FOR{$i = k-m+1: {k}$}
        \STATE $\beta = \rho_i\,\langle\mathbf{y}_i,\mathbf{r}\rangle$
        \STATE $\mathbf{r} = \mathbf{r} + (\alpha_i - \beta)\mathbf{s}_i$
     \ENDFOR
    \STATE \textbf{Return} $\mathbf{\eta}_{k+1}=-\mathbf{r}$. % algorithmic 宏包没有独立的 \Return 命令
\end{algorithmic}
\end{algorithm}

\subsubsection{Riemannian BFGS Method}

Quasi-Newton methods are highly effective for unconstrained optimization as they build an approximation of the Hessian of the objective function. For simplicity, we define $\mathbf{g}_k= \mathrm{grad}\,\tilde{f}_g(\mathbf{\Theta}_{g,k})$. First, the search direction $\mathbf{\eta}_k$ at iteration $k$ is computed by
$\mathbf{\eta}_k = -\mathcal{B}_k\mathbf{g}_k,$
where $\mathcal{B}_k$ approximates the inverse Hessian
matrix and its update is required to satisfy the quasi-Newton condition:
  $\mathcal{B}_{k+1}\mathbf{s}_k=\mathbf{y}_k,$
where $\mathbf{s}_k=\mathcal{T}_{\mathbf{\Theta}_{g,k} \to \mathbf{\Theta}_{g,k+1}}(\alpha_k \boldsymbol{\eta}_k)$ and $\mathbf{y}_k = \mathbf{g}_{k+1} - \mathcal{T}_{\mathbf{\Theta}_{g,k} \to \mathbf{\Theta}_{g,k+1}}(\mathbf{g}_k )$. We denote the transport of a tangent vector $\boldsymbol{\eta} \in T_{\mathbf{x}}\mathcal{M}$ from the tangent space at point $\mathbf{x}$ to the tangent space at point $\mathbf{y}$ as $ \mathcal{T}_{\mathbf{x} \to \mathbf{y}}(\boldsymbol{\eta}) \in T_{\mathbf{y}}\mathcal{M}.$
In the manifold, the update of the variable $\mathcal{B}_k$ is expressed as 
\begin{equation}
\mathcal{B}_{k+1} = \left(\mathbf{I} - \frac{\mathbf{s}_k \mathbf{y}_k^T}{\mathbf{s}_k^T \mathbf{y}_k}\right) \tilde{\mathcal{B}}_k \left(\mathbf{I} - \frac{\mathbf{y}_k \mathbf{s}_k^T}{\mathbf{s}_k^T \mathbf{y}_k}\right) + \frac{\mathbf{s}_k \mathbf{s}_k^T}{\mathbf{s}_k^T \mathbf{y}_k} \label{mathb}
\end{equation}
where $\tilde{\mathcal{B}}_k = \mathcal{T}_{\boldsymbol{\Theta}_{g,k} \to \boldsymbol{\Theta}_{g,k+1}} \circ \mathcal{B}_k \circ \mathcal{T}_{\boldsymbol{\Theta}_{g,k} \to \boldsymbol{\Theta}_{g,k+1}}^{-1}$\cite{huang}.
The above formulas are somewhat different from the established theories of Euclidean geometry. That is because we need to ensure that all vector and operator operations are performed in the same tangent space on the manifold.
However, the R-BFGS method requires the storage and updating of an explicit matrix representation of the operator $\mathcal{B}_k$.
This drawback renders the full R-BFGS method impractical.
\subsubsection{Proposed Riemannian L-BFGS}
To overcome the high complexity of the R-BFGS method, we propose a R-L-BFGS algorithm. The key idea of L-BFGS is to avoid forming, storing, and updating the dense inverse Hessian approximation $\mathcal{B}_k$. Instead, it only stores a limited history of the $m$ most recent displacement and gradient-difference vector pairs, $\{(\mathbf{s}_i, \mathbf{y}_i)\}_{i=k-\hat{m}}^k$, where $\hat{m}=\min\{k,m-1\}$.
Based on the full BFGS update formula similar to (\ref{mathb}), the iterative formula for the L-BFGS method can be derived as:
\begin{equation}
\begin{aligned}
    \mathcal{B}_{k+1} &= {} (\mathbf{V}_{k}^T ···\mathbf{V}_{k-m+1}^T) \mathcal{B}_k^0 (\mathbf{V}_{k-m+1} ··· \mathbf{V}_{k}) + \rho_{k-m+1} \\
    &\times (\mathbf{V}_{k}^T ··· \mathbf{V}_{k-m+2}^T) \mathbf{s}_{k-m+1} \mathbf{s}_{k-m+1}^T (\mathbf{V}_{k-m+2} ··· \mathbf{V}_{k}) \\
    & + \dots  + \rho_k \mathbf{s}_k \mathbf{s}_k^T,
    \label{bk}
\end{aligned}
\end{equation}
where $\rho_i = 1/\langle \mathbf{s}_i, \mathbf{y}_i \rangle$ , $\mathbf{V}_i = (\mathbf{I} - \rho_i \mathbf{y}_i \mathbf{s}_i^T)$ and $\mathcal{B}_k^0$ is initialized as a simple positive-definite matrix. $\langle\cdot,\cdot \rangle$ denotes the Frobenius inner product. By using formula (\ref{bk}), we use a two-loop recursion to calculate the search direction $\mathbf{\eta}_k$ to reduce complexity because this method does not need to store $\mathcal{B}_{k}$. The two-loop recursion is presented in Algorithm \ref{alg:two_loop}. 

\begin{algorithm}
\caption{Proposed R-L-BFGS Optimization Algorithm}
\label{alg:simplified_rlbfgs_upper}
\begin{algorithmic}[1]
   \REQUIRE Problem parameters $\mathbf{f}_k, \mathbf{G}_b, G, \boldsymbol{\rho}, \boldsymbol{\tau}, \mathbf{w}$.
   \ENSURE Optimal matrices $\mathbf{\Theta}_\mathrm{t}^*, \mathbf{\Theta}_\mathrm{r}^*$.
    \FOR{$g = 1 : G$}
        \STATE \textbf{Initialization:} $k= 0$, $\mathcal{B}_0$, empty history buffers $S,Y,P$, iteration number $k_{max}$;
        Compute $\mathbf{\eta}_0=-\mathbf{g}_0$.
        \WHILE{$k \textless k_{max}$ and $\|\mathbf{g}_k\| \ge \epsilon$}
 \STATE Find the step size $\alpha_k$ via Backtracking line search.
            \STATE Compute the new iteration $\mathbf{\Theta}_{g,k+1}=R_{\mathbf{\Theta}_{g,k}}(\alpha_k \boldsymbol{\eta}_k)$.
             \STATE Compute $\mathbf{g}_{k+1}$ using (\ref{27}).
              \STATE Transport all vectors in history buffers $\{S, Y\}$ from $T_{\mathbf{\Theta}_{g,k}}\mathcal{M}$ to $T_{\mathbf{\Theta}_{g,k+1}}\mathcal{M}$\footnotemark.
            \STATE Compute new $\mathbf{s}_k, \mathbf{y}_k$ and
          add $(\mathbf{s}_k, \mathbf{y}_k, \rho_k)$ to buffers $S, Y, P$ based on the Cautious Update Check.
             \STATE Use Algorithm \ref{alg:two_loop} to compute $\boldsymbol{\eta}_{k+1}$.
            \STATE $k = k+1$.
        \ENDWHILE
        \STATE Obtain $\mathbf{\Theta}_g^* = \mathbf{\Theta}_{g,k}$ and extract $\mathbf{\Theta}_{\mathrm{t},g}^*, \mathbf{\Theta}_{\mathrm{r},g}^*$ using (\ref{33}).
    \ENDFOR
    \STATE \textbf{Return} $\boldsymbol{\Theta}_\mathrm{t/r}^\star$.
\end{algorithmic}
\end{algorithm}

\begin{algorithm} 
	\caption{Overall Beamforming Design Framework} 
	\begin{algorithmic}[1] 
	\REQUIRE ~ 
	All channels $\mathbf{h}_{l,k,d}$,$\mathbf{f}_{k}$,$\mathbf{G}_{l}$ where ${\forall l}\in\mathcal{L},k\in\mathcal{K}.$
	\ENSURE ~ 
$\boldsymbol{\Theta}_\mathrm{t}^\star,\boldsymbol{\Theta}_\mathrm{r}^\star,\mathbf{w}^\star.$      
       \STATE {Initialize $\boldsymbol{\Theta}_\mathrm{t/r}= \operatorname{diag}(\theta_{t/r,1}, \dots, \theta_{t/r,M}), 
\theta_{t/r,m} = \frac{1}{\sqrt{2}} e^{j \phi_{t/r,m}},
\phi_{t/r,m} \stackrel{\text{i.i.d.}}{\sim} \mathcal{U}[0,2\pi), \; m=1,\dots,M$; initialize $\mathbf{w}$ using Zero-Forcing (ZF) method.}
       \WHILE{no convergence of sum-SE}
       \STATE Update $\boldsymbol{\tau}$ by (\ref{15}).
       \STATE Update $\boldsymbol{\rho}$ by (\ref{15}).
       \STATE Update $\mathbf{w}$ by solving (\ref{00}).
        
        \STATE Update $\boldsymbol\Theta_{\mathrm{t}}$, $\boldsymbol\Theta_{\mathrm{r}}$ according to {Algorithm} \ref{alg:simplified_rlbfgs_upper}.
         \ENDWHILE
    \STATE \textbf{Return }$\boldsymbol{\Theta}_\mathrm{t/r}^\star,\mathbf{w}^\star$, sum-SE.
	\end{algorithmic}\label{alg:2}
\end{algorithm}

To ensure the robustness and convergence of our R-L-BFGS algorithm in the non-convex problem, we incorporate a Cautious Update Check\cite{huang}. This strategy ensures that the new pair $(\mathbf{s}_k, \mathbf{y}_k)$ is admitted into the history only if it satisfies a more stringent, adaptive curvature condition:
$
    {\langle \mathbf{s}_k, \mathbf{y}_k \rangle_{\mathbf{\Theta}_{g,k+1}}}/{\langle \mathbf{s}_k, \mathbf{s}_k \rangle_{\mathbf{\Theta}_{g,k+1}}} \ge \nu(\|\mathbf{g}_k\|),
    \label{eq:cautious_update}
$
where $\nu(\cdot)$ is a strictly increasing function with $\nu(0)=0$. 
The Cautious Update Check ensures a sufficient descent direction by maintaining a positive-definite Hessian approximation, which, in conjunction with the Armijo condition, guarantees convergence to a stationary point.
At least we obtain reflective and transmissive matrices in each group from $\boldsymbol{\Theta}_g^\star$, i.e.,
\begin{equation}\boldsymbol\Theta_{\mathrm{t},g}^{\star}=[\boldsymbol\Theta_{g}^{\star}]_{1:\bar{M},:},\quad\boldsymbol\Theta_{\mathrm{r},g}^{\star}=[\boldsymbol\Theta_{g}^{\star}]_{\bar{M}+1:2\bar{M},:},\forall g\in\mathcal{G}.\label{33}\end{equation}
The procedure of obtaining $\boldsymbol{\Theta}_\mathrm{t/r}^\star$ is outlined in {Algorithm} \ref{alg:simplified_rlbfgs_upper}. {Algorithm} \ref{alg:2} outlines the overall beamforming design framework, with the complexity being $\mathcal{O}\{I\left(I_{RL}G\bar{M}^3+KLM^2+I_aL^2N^2K^2\right)\}$. Here, $I$, $I_{RL}$, and $I_a$ denote the iterations in Algorithm 3, Algorithm 2, and the PDS method for solving $\mathbf{w}$, respectively.
The computational complexity is 
$\mathcal{O}\{I(I_{RL} M^3 + K L M^2 + I_a L^2 N^2 K^2)\}$ for the FC case ($G$ = 1)
and $\mathcal{O}\{I(I_{RL} M + K L M^2 + I_a L^2 N^2 K^2)\}$ for the SC case ($G$ = $M$).

\footnotetext{The vector transport we choose is isometric, ensuring that the inner product between history vectors remains unchanged during the transport process. Therefore, the scalar value stored in $P$ remains unchanged during the iteration and does not need to be recalculated.}

\begin{figure*}[t]
\centering
\subfloat[]
{\label{figure_ca}\includegraphics[width=0.25\textwidth]{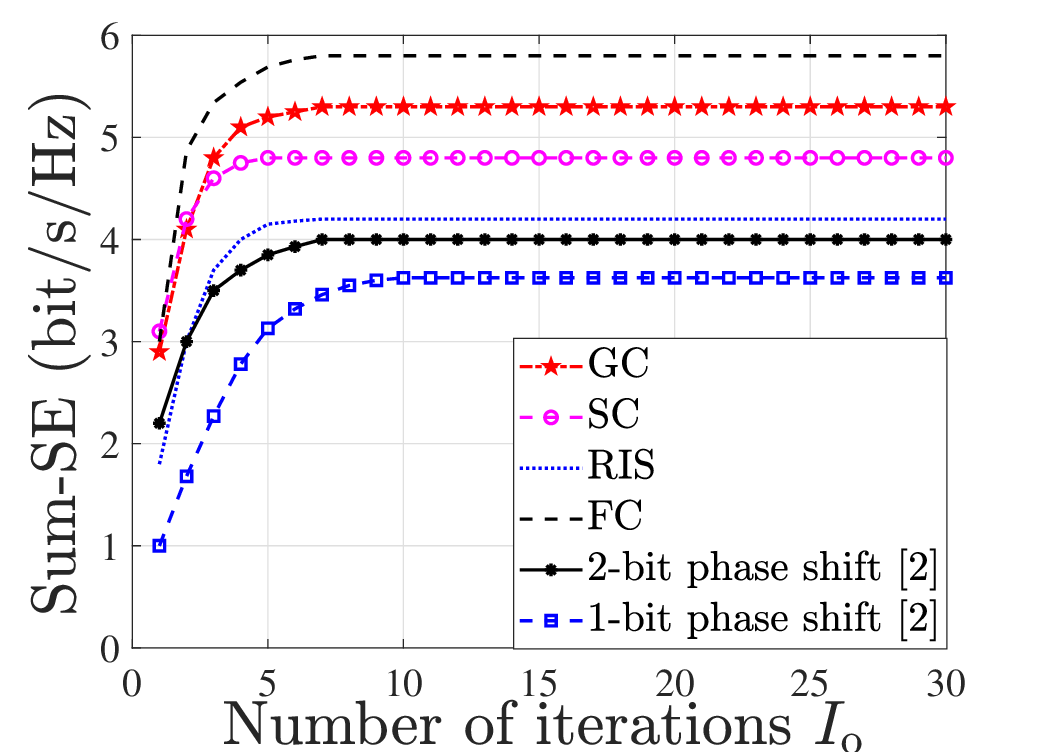}}\hfill
\subfloat[]
{\label{figure_cb}\includegraphics[width=0.25\textwidth]{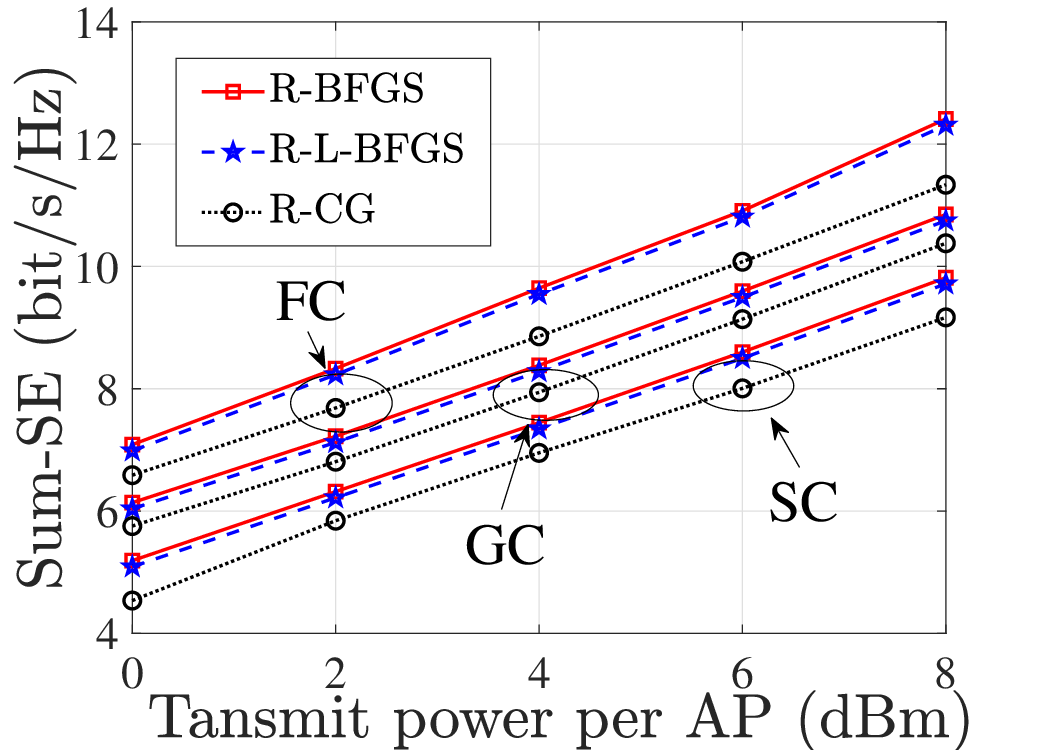}}\hfill
\subfloat[]
{\label{figure_cc}\includegraphics[width=0.25\textwidth]{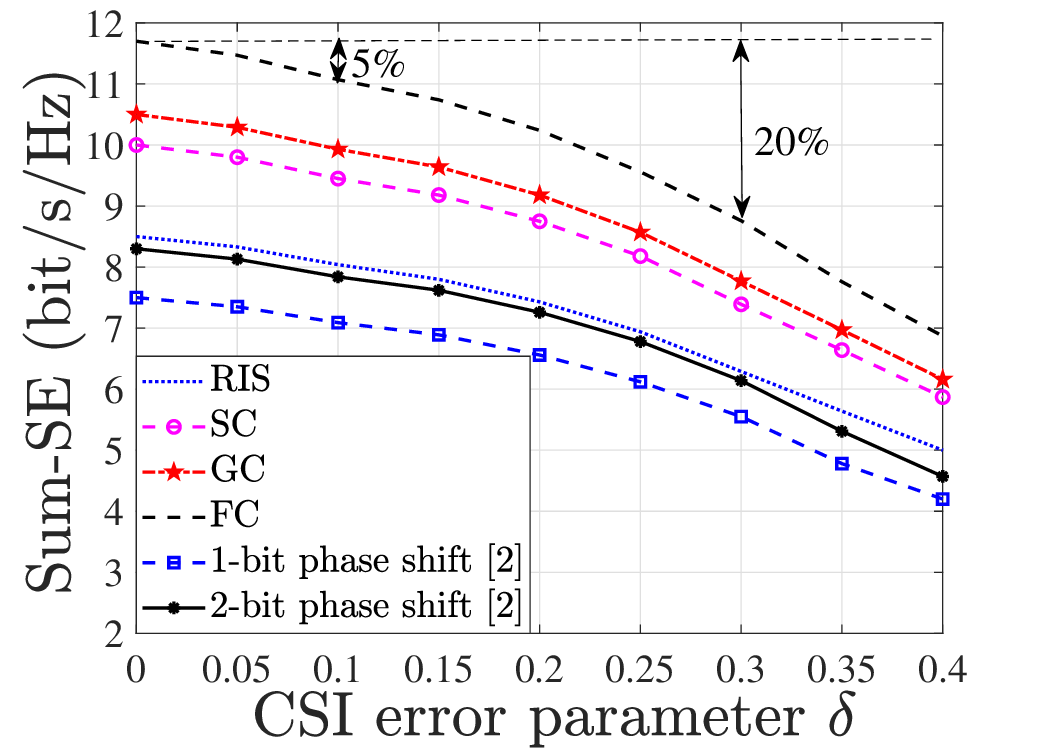}}
\subfloat[]
{\label{figure_cd}\includegraphics[width=0.25\textwidth]{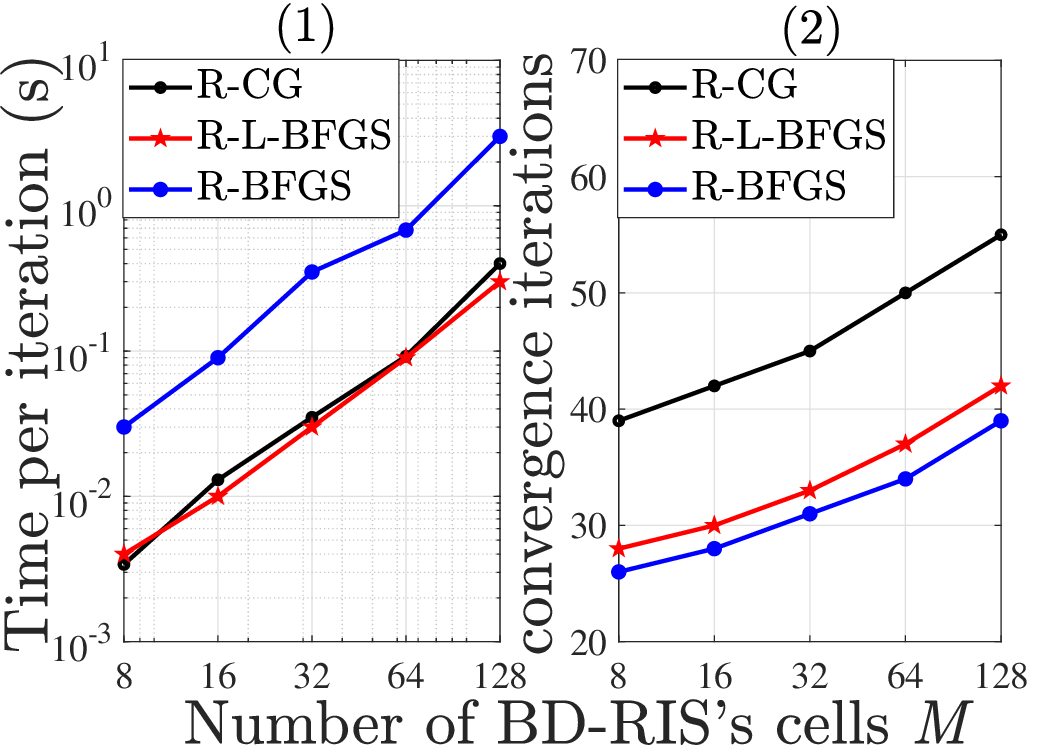}}
\caption{(a): Sum-SE against the number of $\text{iterations } I_{\mathrm{o}}$ with $L=3$, $K=4$, $M=16$, $N=2$, ${P}_{l,\mathrm{max}}=0.001$, $G=2$. (b): Sum-SE against the transmit power per AP with $L=3$, $K=4$, $M=32$, $N=2$, $G=2$. (c): Sum-SE against the CSI error parameter $\delta$ with $L=3$, $K=4$, $M=32$, $N=2$, ${P}_{l,\mathrm{max}}=0.003$, $G=2$. (d): Computational complexity comparison: (1) average CPU time per iteration and (2) average number of iterations to convergence, versus the number of BD-RIS cells $M$. The experimental platform is built on a Windows 11 system with 13th Gen Intel(R) Core(TM) i7-13650HX.}
\vspace{-4mm}
\label{figure_bsf}
\end{figure*}

\section{Simulation Results}
\vspace{-2mm}

This section presents the simulation results to show the performance of the BD-RIS-aided CF mMIMO system and our proposed algorithm. In our system model, the AP–BD-RIS, BD-RIS–UE, and AP–UE links are modeled by both large-scale and small-scale fading\cite{3}. The large-scale fading is modeled by a distance-dependent path loss for a link $i$ with distance $d_i$, given by $PL_i = \zeta_0 (d_i/d_0)^{-\beta}$, where $\zeta_0 = -30\,\mathrm{dB}$ at $d_0 = 1$ m and the path loss exponent $\beta = 2.2$ \cite{1}. The small-scale fading is modeled by a Rician distribution with K-factor $K_{TR}=5$ dB, incorporating both line-of-sight (LoS) and non-line-of-sight (NLoS) components. The LoS channel is represented as the outer product of transmit and receive array response vectors under a ULA assumption, while the NLoS component is modeled by a matrix of i.i.d. complex Gaussian variables\cite{1}. The noise power at each user is set as $\sigma_{k}^{2}=-80 \, \mathrm{dBm},\forall k\in\mathcal{K}$.
And we assume that there are three APs placed at coordinates (0, 0), (0, 10), and (0, -10) meters, and the BD-RIS is placed at (200, 0) meters. The UEs are placed on the left and right halves of a circle centered at the BD-RIS, with two UEs in each half, and a radius of 2.5 meters.
In our simulation, six curves are defined as follows: 1) GC/FC: group-connected/fully-connected BD-RIS;
2) STAR-RIS: single-connected BD-RIS; 3) RIS: traditional diagonal RIS; and 4) 1-bit/2-bit phase shift: non-ideal RIS phase shift case, denoting the cases $\mathcal{F}_{2}$ and $\mathcal{F}_{3}$ in \cite{3}.

Fig. \ref{figure_ca} shows the sum-SE of the system under different architectures in relation to the number of iterations.
We can observe that the proposed algorithm always converges in a limited number of iterations under different BD-RIS architectures, which also demonstrates the robustness of our proposed algorithm.
The result also highlights that BD-RISs can achieve superior sum-SE performance compared to RISs.
We present the sum-SE as a function of the transmit power per AP in Fig.~\ref{figure_cb}. First, regarding the BD-RIS architectures, the performance is consistently ranked as FC {\textgreater} GC {\textgreater} SC, which validates that higher degrees of freedom in the BD-RIS connectivity lead to superior performance. Second, for every architecture, the performance of the R-BFGS algorithm and our proposed R-L-BFGS algorithm are nearly identical and consistently outperform the Riemannian Conjugate-Gradient (R-CG) method. This demonstrates the benefit of leveraging approximate second-order curvature information to converge to higher-quality solutions. The fact that the limited-memory R-L-BFGS can match the performance of the computationally intensive R-BFGS highlights its efficacy for this problem. Fig.~\ref{figure_cc} illustrates the robustness of the proposed beamforming scheme to CSI error. We model the estimated channel $\hat{h}$ as $\hat{h} = h + e$, where $h$ is the real channel and $e$ is the Gaussian estimation error with variance $\sigma_e^2 = \delta |h|^2$ and zero mean\cite{3}. $\delta$ quantifies the CSI error level as the ratio of the error power $\sigma_e^2$ to the channel gain $|h|^2$. As can be observed in Fig. \ref{figure_cc} the sum-SE of all architectures degrades as $\delta$ increases. Particularly, the system performance suffers a loss of 5\% when $\delta$ = 0.1 and a loss of 20\% when $\delta$ = 0.3. Therefore, the proposed algorithm shows strong robustness to CSI error.
Fig.~\ref{figure_cd} provides a comprehensive analysis of the computational performance by comparing the per-iteration CPU time (Fig.~\ref{figure_cd} (1)) and the required convergence iterations (Fig.~\ref{figure_cd} (2) for the R-L-BFGS, R-CG, and R-BFGS algorithms. As shown in Fig.~\ref{figure_cd} (1), the per-iteration cost of R-BFGS grows at an unsustainable rate, confirming its impracticality for large-scale systems, while the costs of R-L-BFGS and R-CG are nearly identical. Fig.~\ref{figure_cd} (2) shows that our proposed R-L-BFGS algorithm requires significantly fewer iterations to converge than the first-order R-CG method, approaching the rapid convergence of the R-BFGS method. For instance, when the number of BD-RIS cells is $M = 32$, the R-L-BFGS algorithm converges in approximately 32 iterations, compared with about 45 iterations for R-CG, yielding a reduction of nearly 28\% in iteration count. Meanwhile, its per-iteration CPU time (around $10^{-2}$ s) remains almost the same as R-CG and is markedly lower than that of R-BFGS (around $10^{-1}$ s). Therefore, our R-L-BFGS algorithm demonstrates an optimal trade-off: it leverages approximate second-order information to reduce the number of iterations, a benefit that far outweighs its minor per-iteration overhead, thus achieving superior overall computational efficiency compared to the R-CG baseline.

\vspace{-3mm}
\section{Conclusion}

In this correspondence, we have investigated the sum-SE maximization problem in a hybrid BD-RIS-aided CF mMIMO system. We proposed a comprehensive beamforming design framework based on alternating optimization. To address the non-convex subproblem, we develop a robust and computationally efficient algorithm based on R-L-BFGS on the Stiefel manifold. Simulation
results have shown that our algorithm can achieve faster convergence and find higher-quality solutions. Therefore, our proposed algorithm has great potential for practical application of hybrid BD-RIS in CF mMIMO systems. Future work of robust beamforming under imperfect CSI and the extension of our framework to novel BD-RISs, such as stem-connected and band-connected architectures, will provide greater theoretical and practical value.

\vspace{-3mm}
\bibliographystyle{IEEEtran}
\bibliography{mybib}

\newpage

\end{document}